# First principles study of electron transport through diarythylene transition metal dichalcogenide molecular switch

A. Ramazani[1], F. Shayeganfar[2], V. Sundararaghavan[3], Nicholas Xuanlai Fang[1,*]

[1]Department of Mechanical Engineering, Massachusetts Institute of Technology, MA 02139, USA.

[2]Department of Civil and Environmental Engineering, Rice University, Houston, TX 77005, USA

[3]Department of Aerospace Engineering, University of Michigan, Ann Arbor MI 48109, USA.

**Abstract:** Computational methods are fast becoming an integral part of nanoelectronics design process. With increasing computational power, electron transport simulation methods such as Non-equilibrium Green's function (NEGF) methods now hold promise in study and design of new electronic devices. Single molecule circuits as optimized device size covers a significant electron transmission, which originated of intrinsic molecular properties. In this study, we study and design a single molecule switch based on a transition metal dichalcogenide (TMD) electrode (molybdenum disulfide ($MoS_2$)) and a photo-chromic molecule. The chosen molecule, Diarylethene, is one of the only few thermally irreversible photochromes. The 1T phase of TMD monolayer has metallic properties and can act as a conducting electrode for these molecular switches. Further, the 1T phase can be functionalized using thiol chemistry, which leads to the formation of covalent $C−S$ bonds that enable further addition of functional photochromic groups to the TMD surface. In this report, we compare and contrast different chemistry and spacer groups with respect to their response as a molecular switch, focusing on the ON/OFF transmission ratio at the Fermi level. We identify promising chemistries for further experimentation. If experimentally realized, these switches are expected to become integral part of various applications including molecular memories, photon detectors and logic devices.

---
[*] Corresponding Author, Email: nicfang@mit.edu

# 1. Introduction

Molecular junction of metal/organic interfaces is predicted to transport charge in nanoscale devices, which can be a good candidate to advance several limitations of semiconductor microfabrication [1–3]. Development of single molecule as building block of switches that allow controlled electron transport is an important aim of nanoelectronics [4]. Devices such as memristors, optical memories and logic devices made from these molecular switches are expected to be extremely energy efficient due to alternate heat transport modes such as quantum tunneling. Photochromic molecules are particularly attractive for molecular switches. External stimuli based on light of specific wavelengths can be used to reversibly and reliably modulate conductance states of such molecules. The molecules are characterized as a reversible color change of a chemical species upon photoirradiation. Although vast numbers of photochromic molecules have been so far synthesized, molecules, which exhibit thermally irreversible photochromic reactivity, are limited within a few families of compounds such as flugicides and diarylethenes.

More recently, Yin et al. [5] designed and optimized the molecular switches by using the phenazine-1,6-dicarboxamides, which their core reduction transforms hydrogen bond acceptor of phenazine core into a hydrogen. Moreover, Maurer et al. [6] reported that the transient ion formed by light-induced isomeration azobenzene act as a metal-supported molecular switches [6]. The light irradiation induced switching of tunable diarylethene molecules from their nonconductive to conductive state, which contacted to gold nanoelectrodes [7]. The practical application of single molecule junctions is to distinguish biomolecules such as DNA and RNA [8]. Ohshiro et al. [9] reported that one can measure the current tunneling of DNA and microRNA sequences and diagnose cancer [10–16].

Diarylethene derivatives are the most promising candidates for applications to opto- electronic switches because of their fatigue resistance, high sensitivity and rapid response. However, one primary challenge in using this effect in a device was in achieving a good coupling of the molecule to the electrode, which implied that either reversibility is lost or the device only operates at extremely low temperatures [17]. A prominent breakthrough was achieved recently when a room temperature reversible stable photoswitch based on a graphene-diarylethene junction was discovered. A drawback of graphene used in this device design is the absence of a band gap and the difficulty to introduce one. This is an issue when coupling these switches within a device to other components based on conventional semiconductors. Two-dimensional transition metal dichalcogenides (TMDs) like molybdenum disulfide ($MoS_2$) have now generated a lot of interest as tunable electrode materials recently due to their unique properties. They are composed of two-dimensional X-T-X sheets stacked on top of one another. TMD monolayers possess high mobility of charge carriers, high thermal stability, and the presence of a distinct band gap. With this material, first field-effect transistors, logical circuits and amplifiers have been manufactured. Large-scale production appears to be feasible.

Manufacturing of TMD based molecular switches has not been analyzed thus far. The primary drawback of $MoS_2$ system had been the difficulty in functionalizing the system due to its inertness. Only recently techniques have emerged to allow covalent bridging of organic molecules to TMDs motivating the present study of a TMD molecular switch. It was recently shown that direct covalent functionalization of the basal planes of $MoS_2$ can be done. Covalent $C-S$ bonds are generated enabling further addition of functional groups to the TMD surface. Using the flexibility of this covalent chemistry we theoretically study tethered chromophore diarylethene derivatives for use in molecular switches.

Past experience shows that connecting an active molecule to electrodes can destroy the reversibility of the molecule. A typical diarylethene molecule with open bond configuration will absorb a photon in the ultraviolet spectrum to switch to a closed-ON state [18, 19]. When visible photons are imposed, the molecule will revert back to the OFF state. Likewise, a diarylethene–graphene switch will go to ON state when UV photons are absorbed. However, the electrode is extremely stable in the ON state due to its conjugated interaction with the graphene electrode and will not revert back to the OFF state when visible photons are imposed. To improve upon the design of electrodes to avoid irreversibility, computer calculations are becoming indispensible [20]. These simulations can be used to identify spacer molecules that would prevent unfavorable interactions of electron states in the active molecule with the electrode. In this study, we will use first principles calculations to uncover rules for how to connect diarylethene molecules to TMD-based electrodes to form reliable switches. We explore two types of functional group additions alongside methyl spacer units that are used to link the active molecule to a TMD monolayer. The aim will be to tune the charge transmission characteristics at the chemical potential (Fermi level).

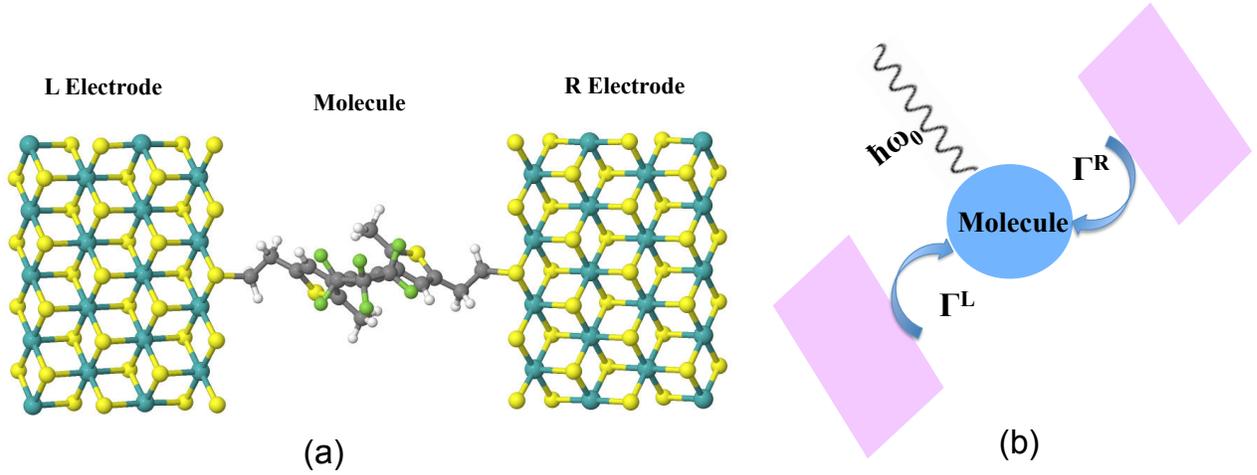

**Fig. 1**: a. Simulation configuration shown the left and right electrodes and the active molecule bridge. b. Schematic representation of one molecule-lead system with coupled leadbound states. The electron-phonon (e-ph) interaction takes place in the molecule region.

## 2. Methodology

The Non-Equilibrium Green's Function formalism (NEGF) is employed for modeling electron transport through nano-scale devices [21, 22]. Electron transport is treated as a one-dimensional coherent scattering process in the scattering region for electrons coming in from the electrodes. The electrodes were made of 1T $MoS_2$ lattice as shown in **Fig. 1**. The active atoms were attached to the sulfur atoms forming a bridge between the two electrodes. When a voltage (V) is applied the Fermi levels shifts as $\mu_L = E_F + eV/2$, $\mu_R = E_F - eV/2$, where $E_F$ is Fermi level of the electrode. The electron states in the molecule between these energy levels assist in electron transport. The transmission function $T(E)$ describes the product of the number (*n*) of available energy states $E$ and the transmission probability of electrons at that energy level. The transmission function is thus critical in describing electron transport through the scattering region. In the NEGF formalism, the transmission function $T(E)$ is computed as:

$$T(E) = Tr[G(E)\Gamma_R(E)G(E)\Gamma_L(E)] \tag{1}$$

where $\Gamma$ is the coupling matrix related to the self-energies of the electrodes and the Green's function $G(E)$ of the system is obtained from solving:

$$(ES - H)G(E) = I \tag{2}$$

where $S$ is the overlap matrix, $H$ is the Hamiltonian and $I$ is the identity matrix. The Hamiltonian is composed as follows ($L$, $C$ and $R$ denote the left lead, the central region and the right lead respectively) and the self-energy ($\Sigma$) of the electrodes (assumed semi–infinite):

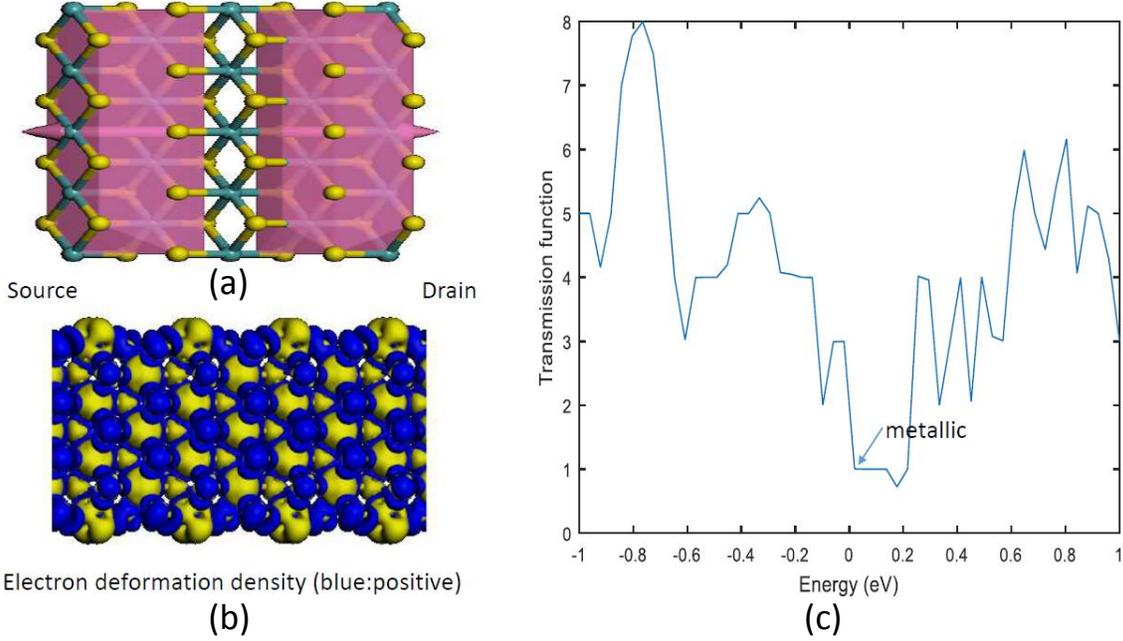

**Fig. 2**: Transmission function of a 1T monolayer electrode and its electron deformation density are shown. The transmission function has a single integer state at Fermi level and is a metallic conductor.

$$H = \begin{bmatrix} H_L + \Sigma_L & H_{LC} & 0 \\ H_{LC} & H_C & H_{RC} \\ 0 & H_{RC} & H_L + \Sigma_R \end{bmatrix} \quad (3)$$

The simulation progresses using three steps. Firstly, the electrodes are modeled with K- space sampling to compute the Hamiltonian matrices $H_L$ and $H_R$ and the Fermi level of the electrode $E_F$. Secondly, a 1D periodic calculation is performed for the scattering region, which includes the leads and the active molecule. This is a gamma point calculation to compute the $H_{LC}$, $H_{RC}$ and $H_C$ Hamiltonian matrices and the corresponding overlap matrices S. The last step includes the calculation of the transmission function T (E) using the above approach, within an energy value of -1 to 1 eV with 50 increments. An example transmission function 1T TMD monolayer electrode is shown in **Fig. 2**. The transmission function has a single integer state at Fermi level and behaves like a metallic conductor as expected.

## 2.1. Electron-Phonon interaction effect on the quantum transport

To investigate the temperature effect on the quantum transport of our devices, we take into account the $e - ph$ interaction as depicted in in **Fig.1**. Its Hamiltonian is written as $H = H_{leads} + H_{ph} + H_{e-ph} + H_M$, where the first term is related to contribution of two leads, which they described as [23]:

$$H_{leads} = \sum_{\alpha k} \epsilon_{\alpha k} C_{\alpha k} C_{\alpha k}^\dagger \qquad (4)$$

where the $C_{\alpha k} C_{\alpha k}^\dagger$ is the creation (annihilation) operator to create (annihilate) an electron in the state $|k>$ for lead ($\alpha =$ L or R) and $E_{\alpha k}$ is the corresponding energy level. The second term is the phonon (*ph*) contribution, which it takes the form as [23]:

$$H_{ph} = \omega_0 a a^\dagger \qquad (5)$$

and $H_{e-ph}$ is the molecule-phonon (M-ph) interaction as:

$$H_{M-ph} = [\epsilon_d + \gamma(a + a^\dagger)] n_d \qquad (6)$$

where, $\gamma$ is the coupling constant between molecule electron and phonon mode, $n_d$ is electron creation/annihilation operators of molecule ($n_d = d^\dagger d$), and $\epsilon_d$ is the single energy level of molecule [23–26]. $H_M$ is the interaction of molecule with leads, which it can be written as:

$$H_M = \sum_{\alpha K} V_{\alpha k} c_{\alpha k}^\dagger d + h.c. \qquad (7)$$

$V_{\alpha k}$ is the coupling between molecule and leads, measuring the overlap of wavefunctions of molecule and leads, and d is the electron annihilation of molecule [27]. To calculate the transmission of molecular device, we use the Green's function, by taking into account the dressed electron-phonon ($e - ph$) interaction. The Hamiltonian with $e - ph$ interaction is changed as [28–32]:

$$H_{e-ph} = \sum_{\alpha k} \epsilon_{\alpha k} C_{\alpha k} C_{\alpha k}^\dagger + \tilde{\epsilon}_d \tilde{d}^\dagger \tilde{d} + \sum_{\alpha k} \tilde{V}_{\alpha k} C_{\alpha k}^\dagger \tilde{d} + h.c. \qquad (8)$$

Where, $\tilde{\epsilon}_d = \epsilon_d - g\omega_0$, and $g = \frac{\gamma^2}{\omega_0^2}$ [28 − 32]; and $\tilde{V}_{\alpha k} = V_{\alpha k} X$ is dressed molecule-lead coupling and X is due to the canonical transformation of particle operator; $X = exp\left[\frac{-\gamma}{\omega_0}(a^\dagger - a)\right]$.

We conclude that the $e - ph$ interaction shifts the molecule-lead coupling. To calculate the localized phonon modes in this study, we replace the expectation value of $X$ instead of its operator, i.e. $<X> = exp[-g(N_{ph} + 1/2)]$. $N_{ph}$ is the phonon population as: $N_{ph} = (exp(\beta\omega_0)) - 1$, where $\beta = \frac{1}{K_B T}$.[23]. In this study, we discuss the influence of two-zero and non-zero temperature on the transmission. The differential conductance for non-zero temperature can be expressed as [23]:

$$G = \frac{\beta e^2}{4h} \sum_{l=-\infty}^{\infty} L_1 \Gamma \tilde{\Gamma} \int d\omega [A_l(\omega)(\tilde{\rho}(\omega - l\omega_0)) + 2B_1(\omega)\tilde{G}_{d,eh}^R(\omega - l\omega_0)] \quad (9)$$

At finite temperature, $L_1$ equals:

$$L_1 = e^{(-g(2N_{ph}+1))} e^{(l\omega_0\beta/2)} J_1(2g\sqrt{(N_{ph}(N_{ph}+1)}) \quad (10)$$

Where $J_1$ is the complex-argument Bessel function. In this formula $A_1(\omega)$ defined as:

$$N_{ph}(N_{ph} + 1)$$

$$\begin{aligned} A_l(\omega) = &\{f_e^L(\omega)[1 - f_e^L(\omega)] + f_e^R(\omega)[1 - f_e^R(\omega)]\}.\{2 \\ &+ [e^{-l\beta\omega_0} - 1][f_e^R(\omega - l\omega_0) + f_e^R(\omega - l\omega_0)]\} \\ &+ [e^{-l\beta\omega_0} - 1][f_e^L(\omega) \\ &- f_e^R(\omega)].\{[f_e^L(\omega - l\omega_0)[1 - f_e^R(\omega - l\omega_0)] - [f_e^R(\omega - l\omega_0)[1 - f_e^R(\omega - l\omega_0]]]\}, \end{aligned}$$

where, $f_e^{L(R)}$ is Fermi distribution of electron in the Left (Right) lead. And $B_l(\omega)$ defined as [33,34]:

$$B_l(\omega) = \{f_e^L(\omega)[1 - f_e^L(\omega)] + f_e^R(\omega)[1 - f_e^R(\omega)]\}.[e^{-l\beta\omega_0} + 1] \quad (11)$$

It is worth to note that the transmission function is: $T_{\alpha\alpha'}(\omega) = Tr[\Gamma_\alpha G^R \Gamma_{\alpha'} G^A]$, where $\Gamma_\alpha$ represents the coupling matrix between the molecule and leads and defined as $\Gamma_\alpha = 2\pi \sum_K |V_{\alpha k}|^2 \delta(\omega \mp \epsilon_{\alpha k})$, and $G^R$ and $G^A$ are the related and advanced Green function [23, 33, 34].

## 3. Results and discussion

The 1T electrode can be functionally bonded to the photochrome using either a $CH_3$ or a $CH_2 - CO - NH_2$ groups as shown in the experimental work in [35] and depicted in **Fig. 3**. This system can then be attached to photochromes. Using this chemistry, we have simulated four different diarylethene molecule attachments both in the OFF and ON states. The eight different configurations that we simulated are shown in **Fig. 4**. All configurations were built and the energy minimized to achieve four primary spacer groups. These are connection via $CH_2$ spacer, $CH_2 - CH_2$ spacer, $CH_2 - CO - NH-$ and $CH_2 - CO - NH - CH_2$ spacergroups respectively.

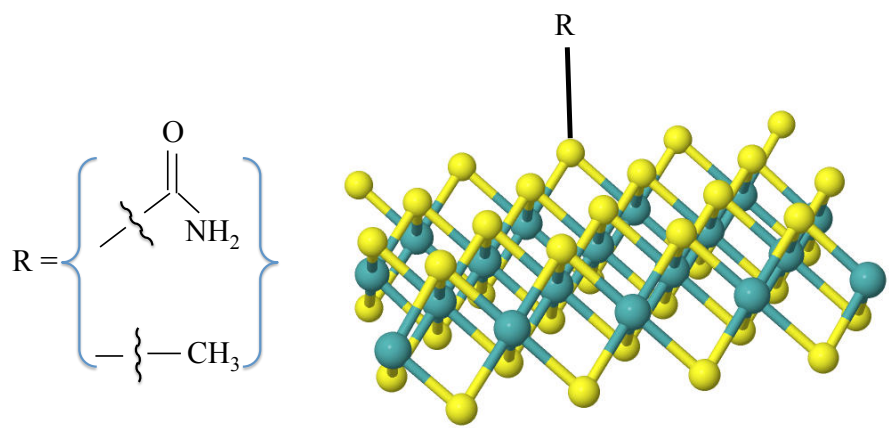

**Fig. 3**: Covalent functionalization of 1T TMDs.

DFT calculations reveal the distribution of negative electron density in the molecule after functionalization. In the OFF state (see **Fig. 5**), the central double bond with a positive electron density is surrounded by unfavorable regions with negative electron density. In the ON position, the conjugated system develops with positive electron density throughout the path from the left and right electrode. This is reflected in the transmission function, where the transmission is found to be at least an order of magnitude higher in the ON state. The DFT calculations were carried out within the VWN parameterization of the local density approximation function. The core electrons were described using semi–core pseudopotentials and a DNP basis set was employed with a global orbital cutoff at 4.4 Angstroms. A self-consistent iteration tolerance of 1E-5 and maximum number of 100 iterations were employedusing the DMol3 program.

The transmission function was subsequently calculated for all eight configurations. The results are shown in **Fig. 6**. All eight systems tend to have significantly lower transmission than the 1T electrode shown in **Fig. 2**, which is expected due to the effect of covalent functionalization. However, $CH_2-CO-NH-CH_2$ spacer group achieves the best OFF response, where the transmission function is lowest. The spacer groups do not as significantly affect the ON response as the OFF response. In effect, the ratio of the ON and OFF response is best achieved by the $CH_2-CO-NH-CH_2$ spacer group.

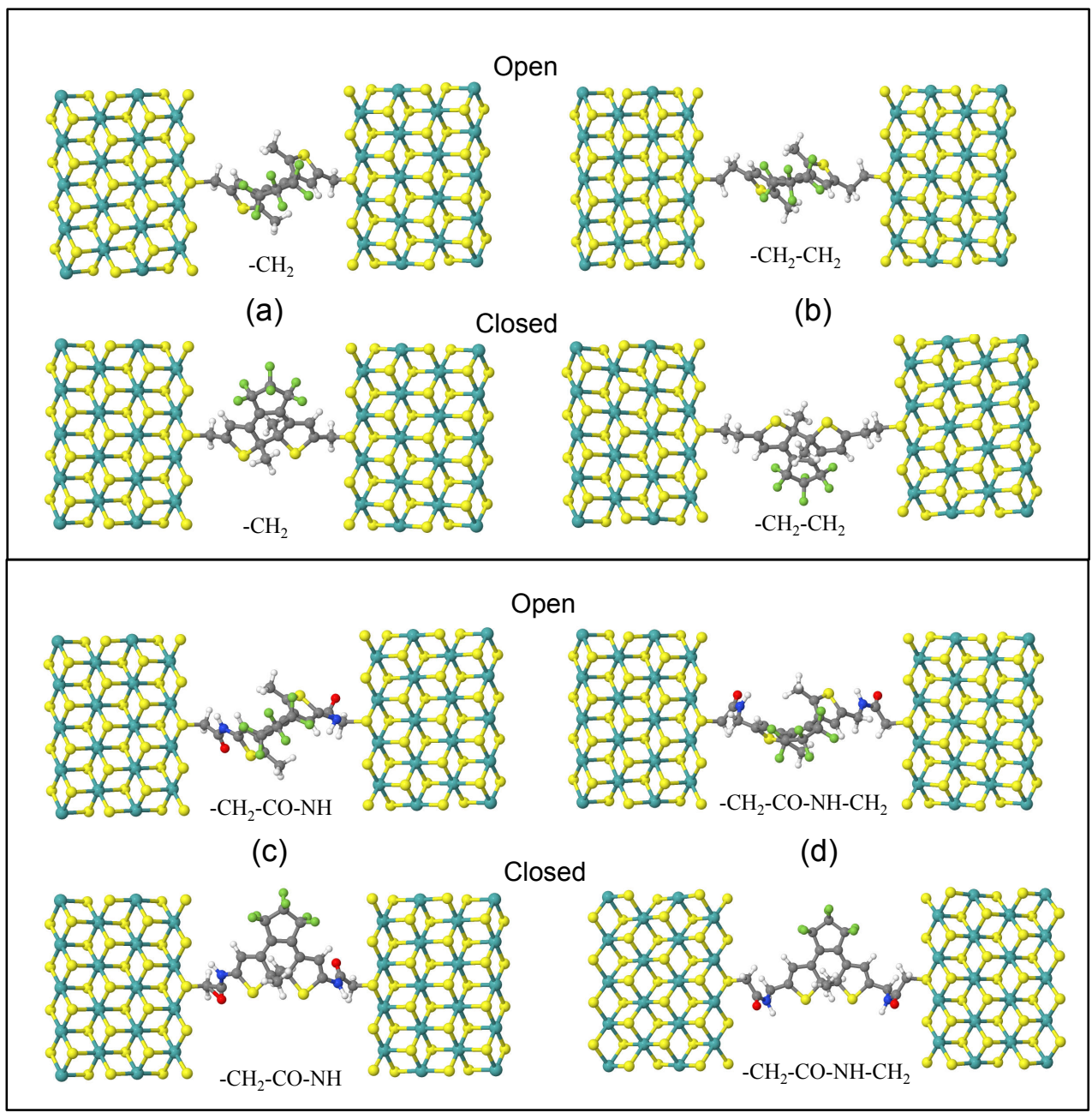

**Fig. 4**: Four primary spacer groups were built at an ON and OFF state. These four switches are based on $CH_2$ spacer, $CH_2-CH_2$ spacer, $CH_2-CO-NH-$ and $CH_2-CO-NH-CH_2$ spacer groups respectively.

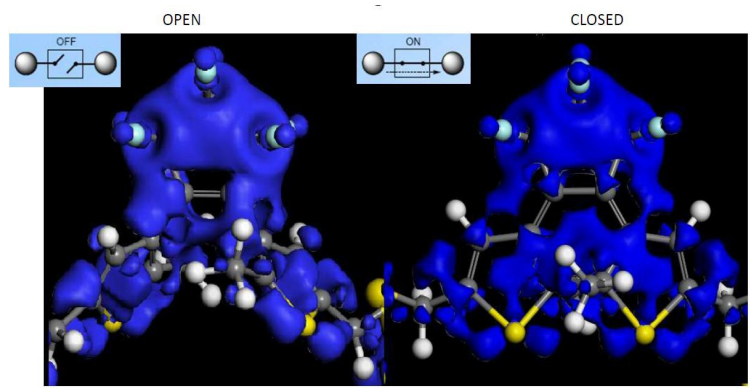

**Fig. 5**: Regions of negative electron deformation density in the ON and OFF states reveals how the conjugated system formed in the ON state mediates an electron flow path from left to right electrode.

Another useful plot to identify the best spacer group is to plot the transmission function at the Fermi level. Understanding of HOMO or LUMO electron states of the molecules close to the Fermi level of the electrodes can be used to well describe the transmission pathway for the electrons [36]. At the Fermi level, the electrons can tunnel to the nearest energy state in the molecule [37]. The transmission function at the Fermi level indicates this tunneling behavior for all eight systems. As seen in **Fig. 6**, the simpler $CH_2$ spacers best mediate the closed transmission. However, the spacers significantly better achieve the OFF state with the $CO-NH$ groups. The ON-OFF ratio is an important aspect of spacer design as this signifies the quality of performance of the switching device and is best achieved the $CH_2-CO-NH-CH_2$ spacer group. Interestingly, this spacer group has even higher ON transmission than the $CH_2-CO-NH$ spacer, which at first look is non–intuitive, but shows the importance of computational analysis in the design of novel molecular switch candidates.

Finally, a series of simulated scanning tunneling microscopy (STM) images were obtained for the different isolated molecular switching and channels with aim of analyzing the charge transport and local states mixing in our systems. **Fig. 9** gives different perspectives of the open/closed molecules, while **Fig. 10** represents the influence of electron transport in the different channels. Both figures show the topographic simulated STM images with a remarkably resolution.

Some features with subtle information are indicated in the simulated STM figures. First, the opened molecular switching in **Fig. 9** reveals several protrusions, which confirm the local electronic states with different colors, while closed molecular switching's show positive states with blue colors. Moreover, in **Fig. 10** electronic transport through closed channels involve uniform local density of states, while opened channels indicate mixing of states with different densities.

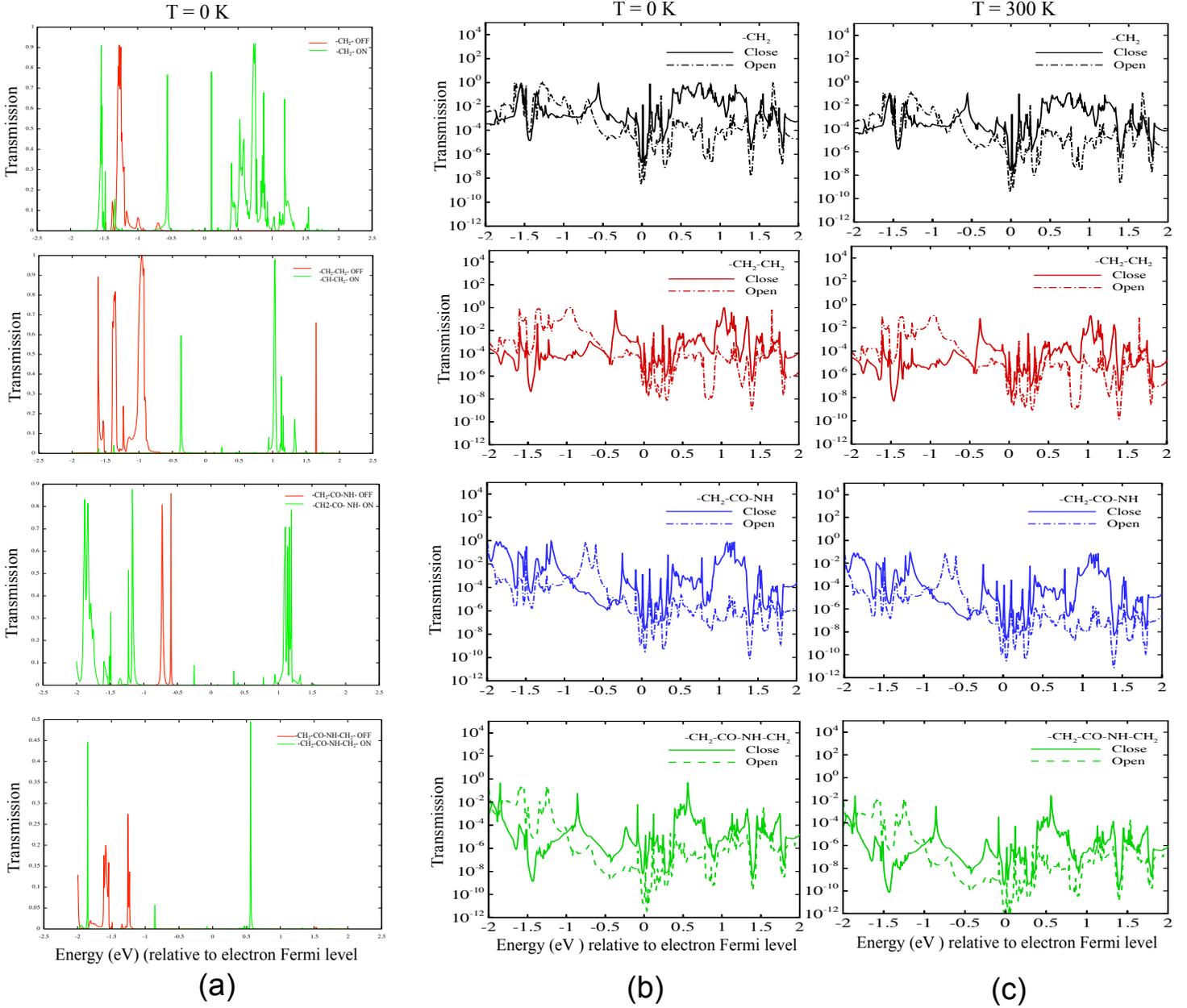

**Fig. 6**: Transmission function for the eight systems at energies relative to the electrode Fermi level at T=0 K and, b. Log-plot of panel a, and c. for T=300 K.

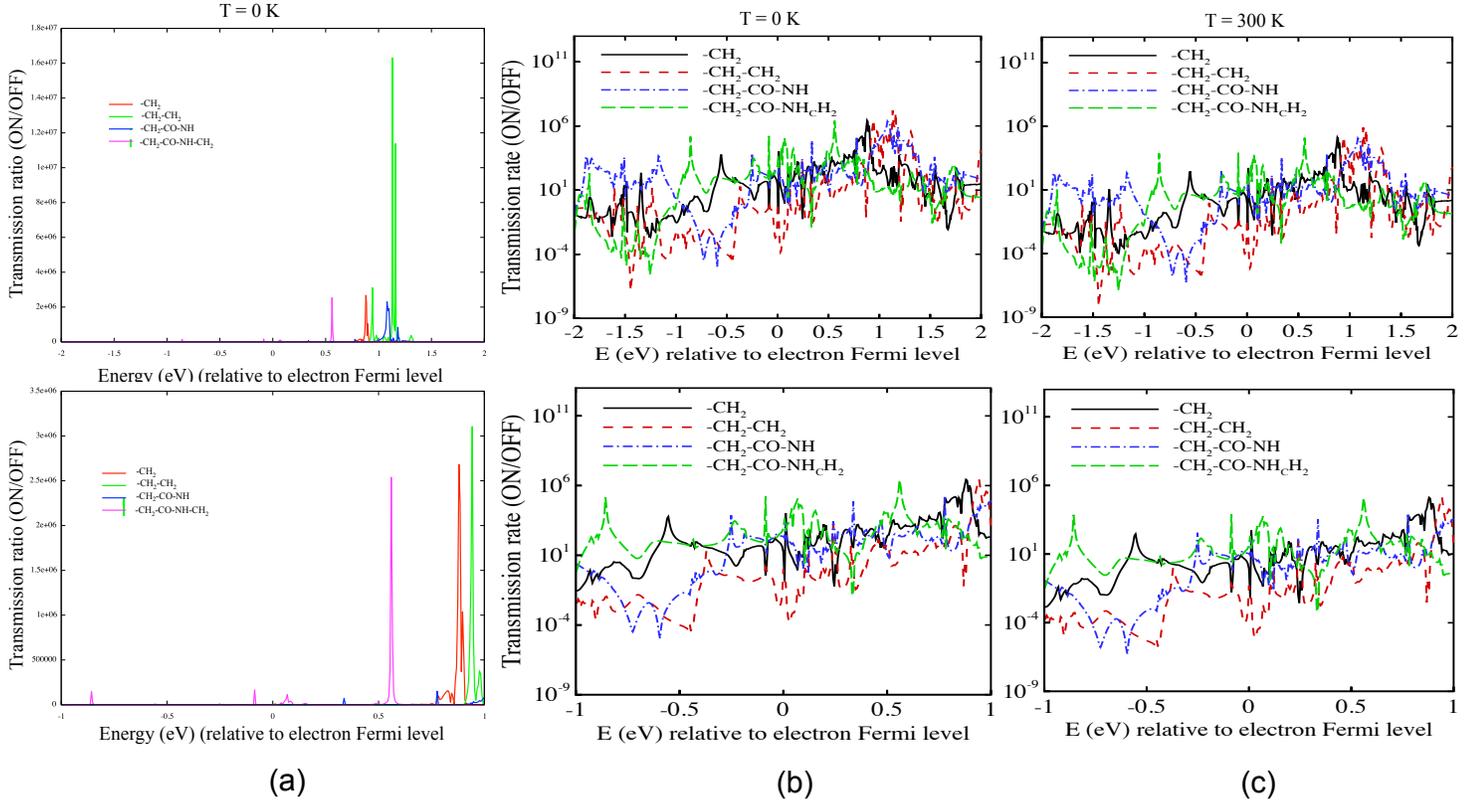

**Fig. 7**: a. The ON/OFF transmission ratio at the Fermi level of the electrodes for all four spacer groups at T=0 K, b. Log-plot of panel a, and c. for T=300 K.

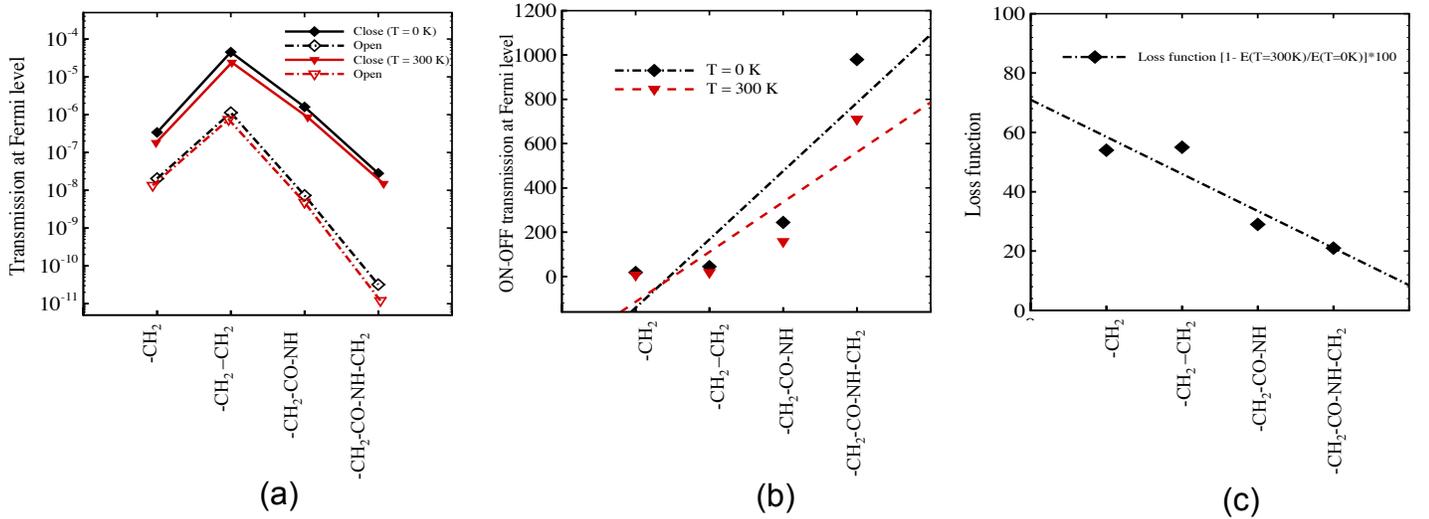

**Fig. 8**: a. The transmission function at the Fermi level for the four systems. b. The ON/OFF transmission ratio at the Fermi level of the electrodes for all four spacer groups and c. The loss function defined as:
[1- E(T=300K)/E(T=0K)] × 100.

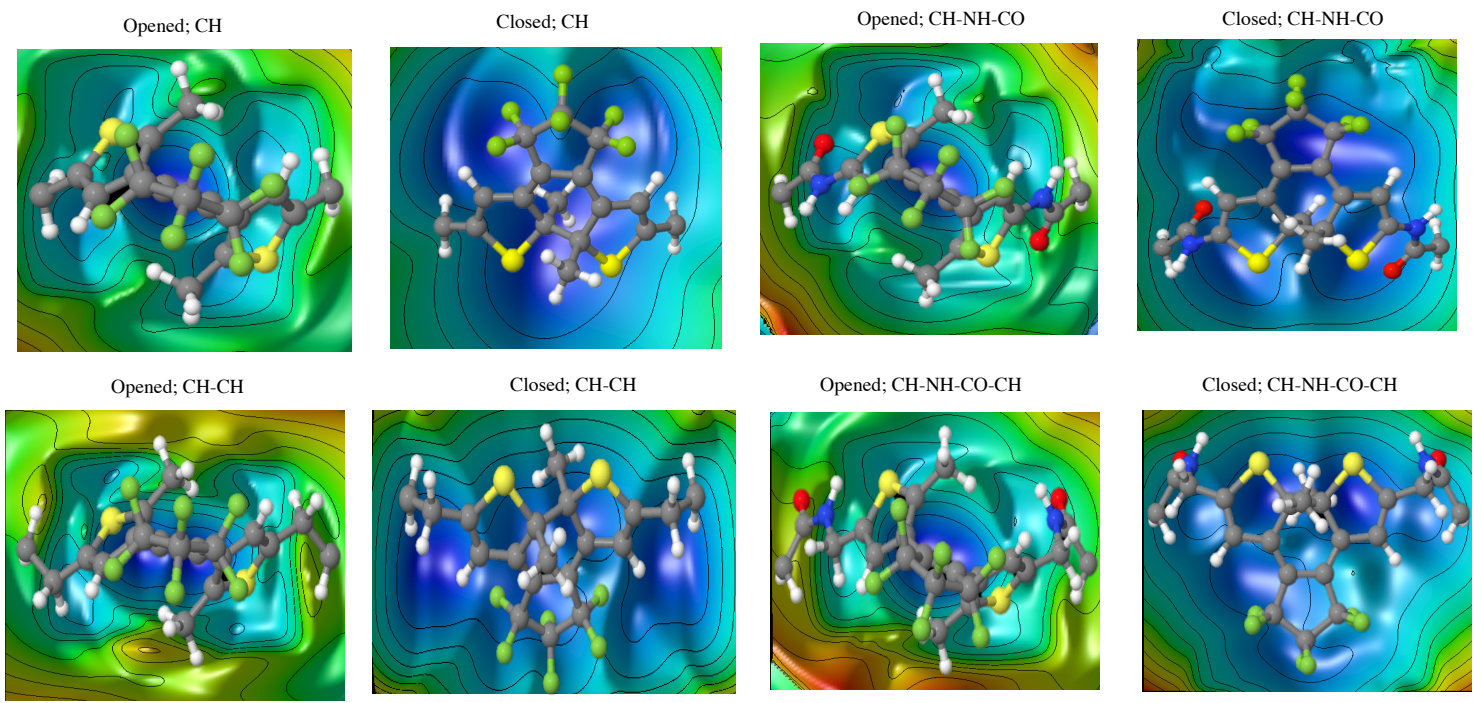

**Fig. 9**: The scanning tunneling microscopy (STM) simulation images for Molecules.

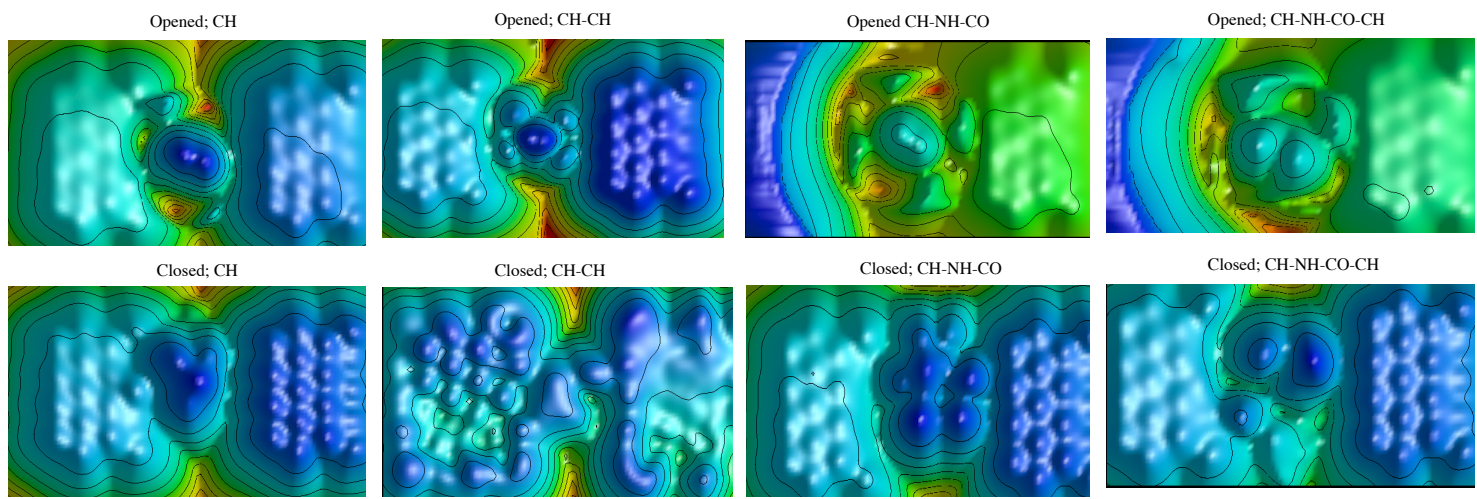

**Fig. 10**: The scanning tunneling microscopy (STM) simulation images for Channels + Molecules.

## Conclusions:

In this study, we designed a single molecule switch based on a transition metal dichalcogenide (TMD) electrode (molybdenum disulfide ($MoS_2$)) and a photo-chromic molecule. The chosen molecule, Diarylethene, is one of the only few thermally irreversible photochromes. The 1T phase of TMD monolayer has metallic properties and can act as a conducting electrode for these molecular switches. Further, the 1T phase can be functionalized using thiol chemistry, which leads to the formation of covalent $C-S$ bonds that enable further addition of functional photochromic groups to the TMD surface. In this report, we compare and contrast four different spacer groups with respect to their response as a molecular switch, focusing on the ON/OFF transmission ratio at the Fermi level. We showed that the ON-OFF ratio is best when using the $CH_2-CO-NH-CH_2$ spacer group. Future work will entail three aspects: 1. understanding the thermal effects through ab-initio molecular dynamics simulations, 2. modeling the current-voltage relationships of the designed system using Landauer formula and 3. understanding the dynamics of electron transport in these devices when cycled through the ON and OFF states. These simulations are valuable to guide experimental realizations of these devices and these switches are expected to become integral part of various applications including molecular memories [38], photon detectorsand logic devices [39].

## Conflict of Interest:

The Authors declare nor conflict of interest.

## Acknowledgements:

The authors would like to thank the U.S. Airforce with the ID No. of …. for supporting this project.